\documentclass[11pt,preprint]{aastex}

\shortauthors{Onken et al.}
\shorttitle{Supermassive Black Holes in AGNs. II.}

\received{}
\accepted{}

\begin{document}

\title{Supermassive Black Holes in Active Galactic Nuclei. II. 
Calibration of the {\boldmath $M_{\rm BH}${\rm --}$\sigma_{\ast}$}
Relationship for AGNs}

\author{ Christopher~A.~Onken\altaffilmark{1},
Laura~Ferrarese\altaffilmark{2}, David~Merritt\altaffilmark{2,3},
Bradley~M.~Peterson\altaffilmark{1}, Richard~W.~Pogge\altaffilmark{1},
Marianne~Vestergaard\altaffilmark{1,4}, and Amri~Wandel\altaffilmark{5} }

\altaffiltext{1}{Department of Astronomy, The Ohio State University, 140
West 18th Avenue, Columbus, OH 43210; onken, peterson,
pogge@astronomy.ohio-state.edu}

\altaffiltext{2}{Department of Physics and Astronomy, Rutgers University,
136 Frelinghuysen Road, Piscataway, NJ 08854; lff@physics.rutgers.edu}

\altaffiltext{3}{Current address: Department of Physics, Rochester Institute of 
Technology, 84 Lomb Memorial Drive, Rochester, NY 14623; drmsps@rit.edu}

\altaffiltext{4}{Current address: Steward Observatory, University of Arizona, 
933 North Cherry Avenue, Tucson, AZ 85721; mvestergaard@as.arizona.edu}

\altaffiltext{5}{Racah Institute, Hebrew University, Jerusalem 91904, Israel;
amri@vms.huji.ac.il}

\begin{abstract}

We calibrate reverberation-based black hole masses in active galactic
nuclei (AGNs) by using the correlation between black hole mass, $M_{\rm
BH}$, and bulge/spheroid stellar velocity dispersion,
$\sigma_{\ast}$. We use new measurements of $\sigma_{\ast}$ for 6 AGNs
and published velocity dispersions for 10 others, in conjunction with
improved reverberation mapping results, to determine the scaling factor
required to bring reverberation-based black hole masses into agreement
with the quiescent galaxy $M_{\rm BH}$--$\sigma_{\ast}$
relationship. The scatter in the AGN BH masses is found to be less than
a factor of 3. The current observational uncertainties preclude use of
the scaling factor to discriminate between broad-line region models.

\end{abstract}


\clearpage

\section{INTRODUCTION}

The advent of techniques for measuring masses of supermassive black holes
(BHs) has led to the identification of correlations between the BH mass
($M_{\rm BH}$) and various properties of the host galaxies. One of the tightest
of these relationships is with the velocity dispersion of the bulge or spheroid
($\sigma_{\ast}$; Ferrarese \& Merritt 2000; Gebhardt et al.\ 2000a). The
objects defining the initial $M_{\rm BH}$--$\sigma_{\ast}$ relationship were
primarily quiescent galaxies, with $M_{\rm BH}$ determined from stellar
kinematics or gas dynamics. However, galaxies hosting an active galactic
nucleus (AGN), in which the BH mass was measured via reverberation mapping
(Blandford \& McKee 1982; Peterson 1993), have been found to be consistent
with following the same correlation (Gebhardt et al.\ 2000b; Ferrarese et
al.\ 2001 [hereafter Paper I]; Onken et al.\ 2003).

The values of $M_{\rm BH}$ derived from reverberation mapping are
subject to certain systematic uncertainties. In tracing the response of
gas in the broad-line region (BLR) to the variable ionizing continuum of
the AGN, the time delay between the fluctuations in the continuum and
the ``reverberation'' of the BLR emission lines gives a characteristic
radius of the BLR gas, $r$. The orbital velocity at that radius is
estimated by the width, $\Delta V$, of the emission line (specifically,
the variable part of the line). But the kinematics and geometry of the
BLR introduces a scaling factor, $f$, into the reverberation mass
equation:
\begin{equation}
M_{\rm BH} = f~\frac{r~\Delta V^2}{G}.
\end{equation}
Simple models of BLR morphologies yield $f$ parameters on the order of
unity, and the application of such estimates, as in the above
references, places the previous AGN $M_{\rm BH}$--$\sigma_{\ast}$ data
among the locus of quiescent galaxies.

A uniform relationship between the BH mass and properties of the host
galaxy on size scales beyond the strong gravitational influence of the
BH implies a causal connection between the formation of the galaxy and
the central black hole. Many investigators have explored possible
mechanisms for how the evolution of the BH and galaxy could be linked
(e.g., Silk \& Rees 1998; Haehnelt \& Kauffmann 2000; Adams, Graff, \&
Richstone 2001; Umemura 2001; Miralda-Escud\'{e} \& Kollmeier 2003;
Merritt \& Poon 2004), while others have looked for outliers from these
relationships as probes of the physical drivers of the correlations
(e.g. Mathur, Kuraszkiewicz, \& Czerny 2001; Wandel 2002; Bian \& Zhao
2004; Grupe \& Mathur 2004).

We present measurements of stellar velocity dispersions for 6
reverberation-mapped AGNs, significantly enlarging the sample of objects
which can be used to observationally investigate the $M_{\rm
BH}$--$\sigma_{\ast}$ relationship for active galaxies. In \S 2, we
describe our observations and analysis method. We present our results
and discuss the implications for the ensemble average value of the
reverberation mapping scaling factor, $\langle f\rangle$, in \S 3. Our
conclusions are summarized in
\S 4.

\section{OBSERVATIONS AND DATA REDUCTION}

The velocity dispersions for our sample of AGNs were measured using the
near-infrared \ion{Ca}{2} triplet (CaT). These stellar absorption lines,
at rest wavelengths of $\lambda$8498, 8542, and 8662~\AA, occur in a region of
relatively low AGN contribution (Nelson \& Whittle 1995), and are
accessible to ground-based observations for sources with redshift
$z\lesssim0.068$ (water vapor bands begin to reduce atmospheric
transparency at longer wavelengths).

Observations were conducted at Kitt Peak National Observatory (KPNO),
Cerro Tololo Inter-American Observatory (CTIO), and MDM Observatory.
The general observing strategy at each of the telescopes was similar.
Long-slit spectra of each target were bracketed by quartz-lamp flat
fields and wavelength calibration exposures. The total exposure time for
each AGN typically exceeded 10,000 seconds. In addition to our targets,
we obtained spectra of late-type giant stars (G8~III--K6~III) to use as
spectral templates. Details of the observing runs are given below and in
Table~\ref{tab1}.

\paragraph{KPNO Observations.}

We observed at the Mayall 4~m telescope with the Ritchey-Chr\'etien (R-C)
Spectrograph. The BL380 grating
(1200 lines mm$^{-1}$, blazed at 9000~\AA) was used with an RG695
blocking filter and the LB1A CCD. The slit width was 2$^{\prime\prime}$, with a
dispersion of 0.45~\AA\ pixel$^{-1}$ and a wavelength range of
8250--9130~\AA.

\paragraph{CTIO Observations.}

We used the V.M.~Blanco 4~m telescope with the
R-C Spectrograph, equipped with the KPGLD-1 grating (790 lines mm$^{-1}$,
blazed at 8500~\AA) and an RG665 filter. We used the lower right
amplifier for the Loral 3K~1 CCD detector within a window of
3072$\times$585 pixels. The reduced spectra have a dispersion of
0.83~\AA\ pixel$^{-1}$ over a wavelength range of 7500--10060~\AA. The
slit dimensions were 2$\times$344$^{\prime\prime}$.

\paragraph{MDM Observations.}

Our first observations at the MDM 2.4~m telescope were made with the MDM
Observatory Modular Spectrograph (ModSpec), using a grating of 830.8
lines mm$^{-1}$, blazed at 8465~\AA. The OG515 order-separating filter
was used. The detector for these observations was ``Charlotte'', a
thinned, backside-illuminated, SITe 1024$\times$1024 CCD.  The spectra
covered a range of $\sim$1400~\AA, with a dispersion of about 1.43~\AA\
pixel$^{-1}$, and a slit width of approximately 2$^{\prime\prime}$.
The second MDM observing run used the ModSpec setup described above, but
utilized a different detector.  ``Echelle'', a thinned,
backside-illuminated, SITe 2048$\times$2048 CCD was used with a
windowing of 330$\times$2048 pixels.

\subsection{Data Reduction and Analysis}

We used IRAF\footnote{IRAF is distributed by the National Optical
Astronomy Observatories, which are operated by the Association of
Universities for Research in Astronomy, Inc., under cooperative
agreement with the National Science Foundation} and XVista\footnote{See
http://ganymede.nmsu.edu/holtz/xvista} for the reduction of different
subsets of the data. However, the basic strategy was the same: the
spectra were flat-fielded, sky-subtracted, and placed on a linear
wavelength scale.

The AGN spectra in our sample were often contaminated by the broad
\ion{O}{1}\ $\lambda$8446 emission line that appears blueward of the 
CaT, and these lines were removed with a high order polynomial in order
to isolate the CaT absorption lines. None of our objects show evidence
of emission from high-order Paschen emission lines, which can also
contaminate the CaT lines.

\section{RESULTS AND DISCUSSION}

\subsection{Measuring Velocity Dispersions}

We measured the velocity dispersions from the reduced spectra with the
Fourier correlation quotient method (FCQ; Bender 1990). Cross-correlating 
the galaxy spectra with those of the template stars produces values for 
the central velocity dispersions.

Our galaxy spectra and the best template-star fits from the FCQ routine are shown in
Figure~\ref{fig1}. The results of our velocity dispersion analysis are
listed in Table~\ref{tab2}. As noted in Paper I, the formal errors reported
by the FCQ fitting routine tend to underestimate the total uncertainties for
the $\sigma_{\ast}$ measurements. Thus, in cases where the FCQ output errors
were smaller than 15\%, the error bars have been brought up to this
threshold.

Based on visual inspection, the least satisfactory fit from
Figure~\ref{fig1} is that of Akn~120. However, consistent (within 1
$\sigma$) results are found from three other methods: (1) a simple
Gaussian fit to the strongest absorption line; (2) the second moment of
the profile for that line; and (3) an IDL routine for penalized pixel
fitting\footnote{See Cappellari \& Emsellem (2004), and
http://www.strw.leidenuniv.nl/$\sim$mcappell/idl/ for details.}.

We examined whether the AGN continuum could dilute the stellar
absorption lines in a manner that would affect the $\sigma_{\ast}$
measurements. This was tested by artificially including an additional
continuum contribution and running the FCQ routine again. Over a large
range of continuum levels, no significant change was seen in the
resulting $\sigma_{\ast}$ values.

Additional stellar velocity dispersion data for galaxies hosting AGNs
have been taken from the literature.

\subsection{Measuring Virial Products}

New analysis of reverberation mapping data has produced updated
measurements of the virial products ($r~\Delta V^2 / G$) for a sample of
35 AGNs, including the 16 which now have velocity dispersions
(Peterson et al.\ 2004). These virial products, which, from equation (1),
can also be written as $M_{\rm BH}/f$, differ from the majority of
previous reverberation mapping analyses in the calculation of $\Delta
V$. Whereas earlier work typically characterized the velocity width via the
full-width at half-maximum (FWHM) of the emission line, Peterson et al.\
(2004) have found more consistent virial products when using the line
dispersion (i.e., the second moment of the line profile; $\sigma_{\rm
line}$) instead. This will have important implications for comparing our
results with previous publications.

\subsection{The \boldmath{$M_{\rm BH}$--$\sigma_{\ast}$} Relation}

We assume that the $M_{\rm BH}$--$\sigma_{\ast}$ correlation found in
quiescent galaxies also holds for AGNs and their host galaxies. We
can then determine the factor required to scale the AGN virial products
($M_{\rm BH}/f$) to the quiescent galaxy relation (holding
$\sigma_{\ast}$ fixed at the measured value); that is, we calculate the
average scaling factor, $\langle f\rangle$, that, when multiplied by
the virial products, brings the AGN $M_{\rm BH}$--$\sigma_{\ast}$
relation into agreement with the quiescent galaxy relationship.

Because published determinations of the slope of the quiescent galaxy
relation have yet to converge, we selected the most prominent slope
values from either end of the quoted range. Thus, we consider both the
Tremaine et al.\ (2002; hereafter T02) slope of 4.02, and the Ferrarese
(2002; F02, henceforth) slope of 4.58, and determine separate scaling
factors, $\langle f_{\rm T}\rangle$ and $\langle f_{\rm F}\rangle$,
respectively.

While questions remain as to the reliability of the current fits to the
quiescent galaxy $M_{\rm BH}$--$\sigma_{\ast}$ relation, especially with
respect to the accuracy of the black hole masses from stellar kinematics
(see Valluri, Merritt, \& Emsellem 2004; Cretton \& Emsellem 2004;
Richstone et al.\ 2004), we hope our use of the two slope values will
give a more reliable estimate of the uncertainties involved while also
avoiding entanglement in the continuing controversy.

The best fit to the AGN $M_{\rm BH}$--$\sigma_{\ast}$ values was
accomplished with the orthogonal regression program,
GaussFit\footnote{GaussFit is available at
ftp://clyde.as.utexas.edu/pub/gaussfit/} (version 3.53; Jefferys,
Fitzpatrick, \& McArthur 1988), which accounts for errors in both
($M_{\rm BH}/f$) and $\sigma_{\ast}$. For the purposes of our GaussFit
analysis, the asymmetric variances in the virial products were
symmetrized as the mean of the upper and lower variances. A fit was made
to the equation
\begin{equation}
\log\left (\frac{M_{\rm BH}}{f}\right ) = \alpha + \beta\ \log\left (\frac{\sigma_{\ast}}{200} \right )\ ,
\end{equation}
where $\alpha$ is the normalization of the relationship, $\beta$ is the
(fixed) value of the slope, and where ($M_{\rm BH}/f$) is in solar
masses and $\sigma_{\ast}$ is in km s$^{-1}$. For each slope, the
best-fit value of $\alpha$ was calculated. The $\chi^{2}$ which was
minimized to determine the best fit is given by
\begin{equation}
\chi^2 \equiv \sum_{i=1}^{N} \frac{\left [(\frac{M_{\rm BH}}{f})_i - \alpha - \beta (\frac{\sigma_{\ast}}{200})_i \right ]^2}
{\sigma_{Mi}^2 + \beta^{\,2} \sigma_{\sigma i}^2}\ ,
\end{equation}
where $\sigma_{{\rm M} i}$ and $\sigma_{\sigma i}$ are the uncertainties
in the virial product and the velocity dispersion, respectively, for the
$i$th data pair (Press et al.\ 1992).  The virial products and velocity
dispersions used in our fits are given in Table~\ref{tab3}.

The published values of the quiescent galaxy intercept correspond to
$\alpha_F$ = 8.22$\pm$0.08 and $\alpha_T$~=~8.13$\pm$0.06. Our fits with
fixed slopes are shown in Figure~\ref{fig2}, and yield $\alpha_F^{AGN}$
= 7.48$\pm$0.13 and $\alpha_T^{AGN}$ = 7.39$\pm$0.12. Remarkably, both
slopes give the same best fit value, $\log \langle f\rangle$ =
0.74. Hence, $\langle f_{\rm F}\rangle$ = 5.5$\pm$1.9 and $\langle
f_{\rm T}\rangle$ = 5.5$\pm$1.7. The $\chi^2$ per degree of freedom,
$\chi^2_{\nu}$, for these two fits are 2.90 and 2.87, respectively.

Table~\ref{tab4} shows the resulting $\alpha$ and $\beta$ values for a
variety of fitting constraints. When we allow $\beta$ to vary in the fit
to the AGNs, the result is consistent with both the T02 and F02 slopes,
albeit with a large uncertainty. The distribution of points with
previous $\sigma_{\ast}$ measurements in Figure~\ref{fig2} differs from
that of Paper I because of the improved virial product data from Peterson
et al.\ (2004).

Because the reverberation mapping results for IC~4329A and NGC~4593
provide only upper limits on the black hole masses, these data were
omitted from the above analysis. The formal error bars on $M_{\rm BH}$ can
extend below zero because the uncertainty of the time delay is not
restricted to positive values. If we take the uncertainty in $\log
M_{\rm BH}$ to be $\sigma_{\rm M}/(M \ln 10)$, we can avoid taking the
logarithms of negative numbers, and these two data points can be
included in the fit. The uncertainties are large enough, however, that
the fit is not changed in a statistically significant way. Additionally,
the fit is not strongly affected by the exclusion of NGC~4051 (see
Table~\ref{tab4}), which seems to be an outlier in the AGN
radius-luminosity relationship (see the Appendix of Vestergaard
2002). When we do not include NGC~4051, we find $\langle f_{\rm
F}\rangle$=5.1 and $\langle f_{\rm T}\rangle$=5.2,
differences of less than 8\% from our best fit value of 5.5.

The most common assumption for converting a virial product to $M_{\rm BH}$ is
that of Netzer (1990)
\begin{equation}
M_{\rm BH} = \frac{3~r~V_{\rm FWHM}^2}{4~G} ,
\end{equation}
which implicitly assumes an isotropic velocity distribution and $\Delta
V$ = $V_{\rm FWHM}~/~2$, where $V_{\rm FWHM}$ is the line full width at
half maximum. Peterson et al.\ (2004) measure the second moment of the
line profile, $\sigma_{\rm line}$, rather than $V_{\rm FWHM}$, so an
isotropic velocity field simply gives $f$ = 3 with no other assumptions.
Denoting the scale factor as $\epsilon$ when using $V_{\rm FWHM}$ and
$f$ when using $\sigma_{\rm line}$, we find that whereas Netzer's
assumptions give $\epsilon$ = 0.75 (as in eq. [4]), the value of
$\langle f\rangle$ we derive above implies $\langle \epsilon \rangle$ =
1.4, i.e., black hole masses $\sim$1.8 times higher\footnote{Our value
of $\epsilon$ is calculated assuming $\sigma_{\rm line}$ = $V_{\rm
FWHM}/2$; for typical line profiles, changing the shape of the line
alters the $\sigma_{\rm line}$--$V_{\rm FWHM}$ conversion by an amount
smaller than the observed scatter in the $M_{\rm BH}$--$\sigma_{\ast}$
data.}.  Our masses are only 62\% as large as those expected from the
$\sigma_{\rm line}$--$V_{\rm FWHM}$ relation found in a model-dependent
fit to the distribution of $V_{\rm FWHM}$ in a sample of quasars and
Seyfert galaxies (McLure \& Dunlop 2001).

Applying our derived value of $\langle f\rangle$ = 5.5 (or $\langle
\epsilon \rangle$ = 1.4) will remove the systematic bias between the virial
product and mass of the black hole (modulo the accuracy of the
particular $M_{\rm BH}$--$\sigma_{\ast}$ fit chosen). Yet, estimates for
individual AGN black hole masses may fall substantially off the $M_{\rm
BH}$--$\sigma_{\ast}$ relation, as there are still rms deviations from
the fits in the mass direction of factors of 2.9 (F02 slope) and 2.6
(T02 slope).

Whether the scatter in $M_{\rm BH}$ is due to differences in the BLR
properties in each AGN (corresponding to slightly different $f$ values
for each object) or a problem in our assumption that all AGNs fall on
the quiescent galaxy relationship remains unclear. However, models of
the BLR may shed light on this question.

\subsection{BLR Models}

Given our determination of $\langle f\rangle$, we can try to learn
something about the structure and kinematics of the BLR.  Modeling of
the BLR, taking into account different radial profiles and discrete time
sampling of example light curves, will be undertaken in a later
paper. Here we describe only a thin, rotating ring. In this case, $f$
can be thought of as relating the Keplerian velocity ($V_{\rm rot}$) at
the radius of the ring (which is traced by the appropriate time delay)
to the observed line dispersion:
\begin{equation}
f = \left (\frac{V_{\rm rot}}{\sigma_{\rm line}}\right )^2,
\end{equation}
For a ring of inclination $i$ (where $i$=$0^\circ$ for face-on), this
model gives $f$ = $2 \ln2~/ \sin ^2 i$. 

Working within a thin-ring model context, Wu \& Han (2001) calculated the
inclinations necessary to scale published reverberation mapping results
for 11 AGNs to the quiescent galaxy $M_{\rm BH}$--$\sigma_{\ast}$
relation. With our revised virial product results, we find that this
very simple model fails for two of our 16 AGNs. NGC~3783 and Mrk~110
would require $\sin i$ values greater than unity to scale them to the
quiescent galaxy $M_{\rm BH}$--$\sigma_{\ast}$ relation. For our other
sources, we do not find any evidence to support the inclination trends
with line width or radio loudness that were claimed by Wu \& Han.
Further constraints on the BLR geometry are not feasible given the sizes
of the current uncertainties.

\subsection{Gravitational Redshifts}

In principle, an independent measure of the AGN black hole mass can be
obtained by detection of gravitational redshifting of the emission lines
(e.g., Netzer 1977; Peterson et al.\ 1984).  Kollatschny (2003b) finds
that the variable parts of the strong broad lines in the optical
spectrum of Mrk~110 are redshifted relative to the systemic velocity.
The reverberation time lag and redshift of each line relative to
systemic are anticorrelated and appear to be consistent with a
gravitational redshift caused by a central mass $M_{\rm grav} \approx
1.4\times10^8\,M_{\odot}$, rather higher than our reverberation-based
mass\footnote{Kollatschny (2003a) obtains a reverberation-based mass
estimate of $1.8\times 10^7\,M_{\odot}$, but using FWHM as the
line-width measure and $\epsilon$ = 1.5 (see \S 3.1). The FWHM data of
Peterson et al.\ (2004) gives a result consistent with that of
Kollatschny, however our preferred method of measuring $\sigma_{\rm
line}$ and using the empirical calibration of $\langle f\rangle$ yields
a black hole mass approximately 40\% larger.} (Table 3) of $M_{\rm BH}$ =
$2.5\times 10^7\,M_{\odot}$.

Regardless of the particular value of $f$ appropriate to Mrk~110, the
offset between the reverberation-based virial product and the
gravitational redshift mass is in the direction expected. However, the
value of $M_{\rm grav}$ places Mrk~110 even further above the $M_{\rm
BH}$--$\sigma_{\ast}$ relation, which predicts 
$M_{\rm BH}~\approx~4\times 10^6\,M_{\odot}$. 

If the BLR is modeled as a simple disk, then
\begin{equation}
\frac{(M_{\rm BH}/f)}{M_{\rm grav}} = \frac{\sin^2 i}{2 \ln 2},
\end{equation}
and the difference between these two measurements allows us to infer
that $i \approx 30^\circ$, slightly larger than the $21 \pm 5$ degrees
found by Kollatschny (2003b). 

Measuring gravitational redshifts appears to be a promising technique
for independently estimating central masses. However, applying the
method is non-trivial because the gravitational redshifts are small,
typically only several tens to a few hundred kilometers per
second. Moreover, it is not a shortcut around reverberation mapping
because (a) the redshift must be measured in the variable part of the
emission line (i.e., the root-mean-square spectrum formed from the
monitoring data; see Peterson et al.\ 2004) and (b) the time lag, which
can vary with time, needs to be measured simultaneously. The spectral
resolution required to measure such small redshifts reliably is higher
than usually employed in reverberation mapping campaigns, with the
exception of Kollatschny's (2003a) program on Mrk~110 and the {\it
Hubble Space Telescope} monitoring program on NGC~5548 (Korista et al.\
1995). A re-examination of the NGC~5548 spectra from the latter campaign
shows some evidence for redshifts at the expected levels, although the
errors are quite large.  Additional data will thus be required to
examine this method of black hole mass measurement more thoroughly and
to determine whether Mrk~110 is truly a significant outlier from the
$M_{\rm BH}$--$\sigma_{\ast}$ relation.

\subsection{\boldmath{$\sigma_{\ast}$} Versus FWHM([\ion{O}{3}]$\lambda$5007 \AA)}

Several recent studies have suggested the use of the FWHM of the
[\ion{O}{3}]$\lambda$5007 \AA\ emission line as a proxy
for $\sigma_{\ast}$. Nelson \& Whittle (1996) examined a sample of 66
Seyfert galaxies with both $\sigma_{\ast}$ and FWHM([\ion{O}{3}])
measurements and found scatter of 0.20 dex around a 1:1 correspondence
(see their Figures 5 and 7a). Based on these trends, Nelson (2000) and
Shields et al.\ (2003) used [\ion{O}{3}] data as a substitute for
$\sigma_{\ast}$ in AGNs out to z$\sim$3. Boroson (2003) applied the same
arguments to data from the Sloan Digital Sky Survey Early Data Release
and found that using FWHM([\ion{O}{3}]) could reproduce an $M_{\rm
BH}$--$\sigma_{\ast}$-like relationship, although with a larger scatter
than has been found for quiescent galaxies (i.e., a factor of
$\sim$5). Statistically identical distributions in FWHM([\ion{O}{3}])
were found for the broad-line and narrow-line Seyfert 1s in an X-ray
selected sample of AGNs, which placed the narrow-line Seyfert 1s
preferentially below the $M_{\rm BH}$--$\sigma_{\ast}$ relation (Grupe
\& Mathur 2004).

With the FWHM([\ion{O}{3}]) data tabulated by Nelson (2000) for our 16
AGNs, we looked at the relationship with $\sigma_{\ast}$. 
Figure~\ref{fig3} shows that 25\% of the sources have 
FWHM([\ion{O}{3}]) data that deviate by $>$0.2 dex from the values
expected based on their velocity dispersions. There is no evidence of
the correlation between the discrepant objects and their radio power
that was found for a larger sample of Seyferts (Nelson \& Whittle 1996),
indicating that these differences may not be due to a systematic acceleration of the
[\ion{O}{3}]-emitting gas by the radio source. Overall, the weighted
mean difference from equality is 0.03 dex (in the sense of larger FWHM
relative to $\sigma_{\ast}$) with an rms scatter of 0.15 dex. In the way
of commentary, we simply echo sentiments expressed by others that while
use of such a proxy may be valid for a large sample of AGNs, it can also fail
dramatically for individual objects.

\section{CONCLUSION}

With the addition of 6 new velocity dispersion measurements for
reverberation-mapped AGNs and making use of improved
reverberation mapping results, we tie together the $M_{\rm
BH}$--$\sigma_{\ast}$ relationships for quiescent galaxies and
AGNs. This allows us to calculate the average scaling factor, $\langle
f\rangle$, which removes the statistical bias between the virial product
generated by reverberation mapping and the black hole mass. For the F02
and T02 fits to the quiescent galaxy $M_{\rm BH}$--$\sigma_{\ast}$
relationship, we find $\langle f\rangle$ = 5.5$\pm$1.9 and 5.5$\pm$1.7,
respectively. These values of $\langle f\rangle$ apply specifically to
virial products using the dispersion of the emission lines, rather than
measurements of the FWHM. While modeling of the BLR and 
studies of emission line gravitational redshifts may eventually lead
to a better understanding of the structure and kinematics of the BLR,
further work is needed.

\acknowledgments

We acknowledge the anonymous referee for comments that helped to improve
the clarity of our manuscript. We thank Scott Tremaine for useful
correspondence, and Michele Cappellari for making his IDL routines
publicly available. We also thank the staffs at MDM, CTIO, and KPNO for
their assistance. We acknowledge support for this work through NSF Grant
AST0205964 to The Ohio State University. A.~W. acknowledges support by
BSF under grant number 1999-336.

This research has made use of the NASA/IPAC Extragalactic
Database (NED) which is operated by the Jet Propulsion Laboratory,
California Institute of Technology, under contract with the National
Aeronautics and Space Administration.

\clearpage

%figures
\plotfiddle{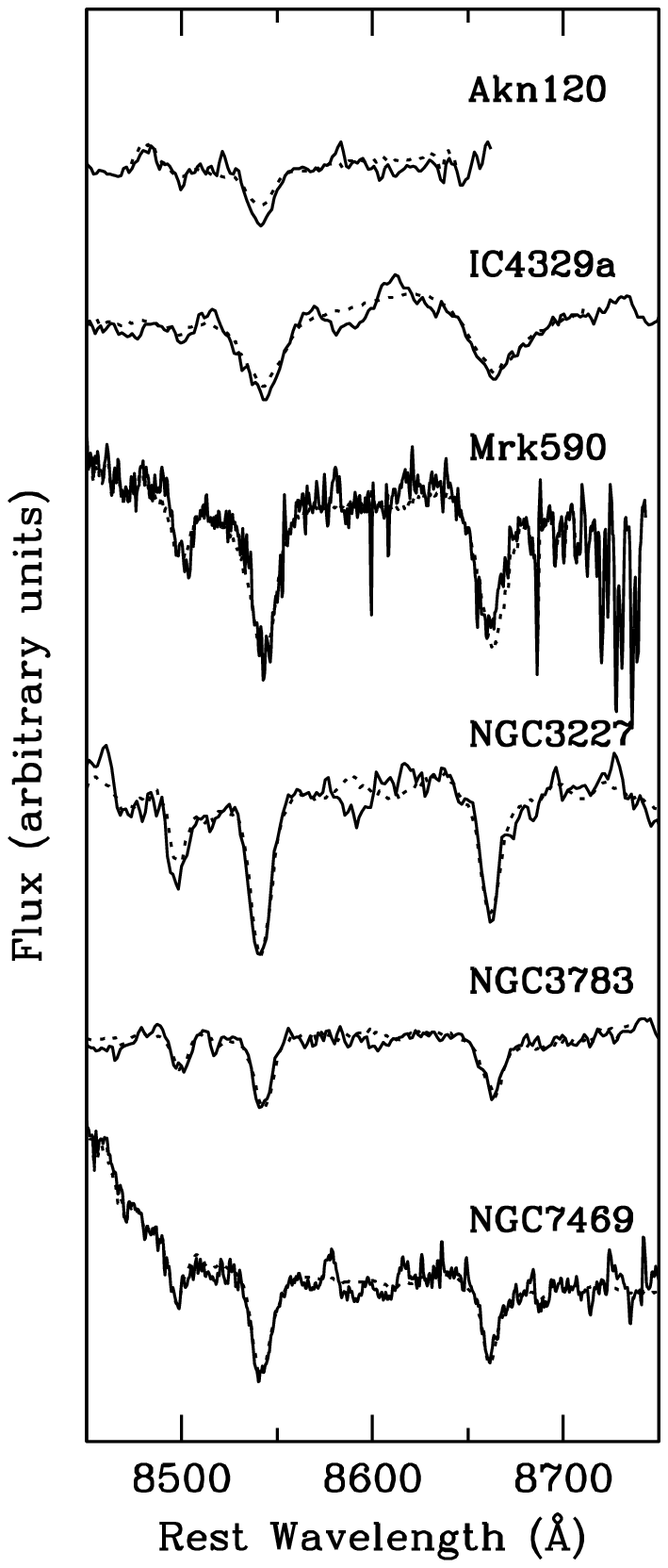}{0in}{0}{250}{500}{0}{0}
%\plotone{f1.eps}
\figcaption[f1.eps]{Normalized spectra of the CaT region for the 6 AGNs in
this study. Each spectrum has been offset in flux for clarity. The
dashed line indicates the best fit obtained with the FCQ method. The
spectrum of Akn~120 is truncated where the FCQ fit becomes unstable. The
spectra shown for Mrk~590 and NGC~7469 are from KPNO; IC~4329A
and NGC~3783 are from MDM. \label{fig1}}

\clearpage

\plotone{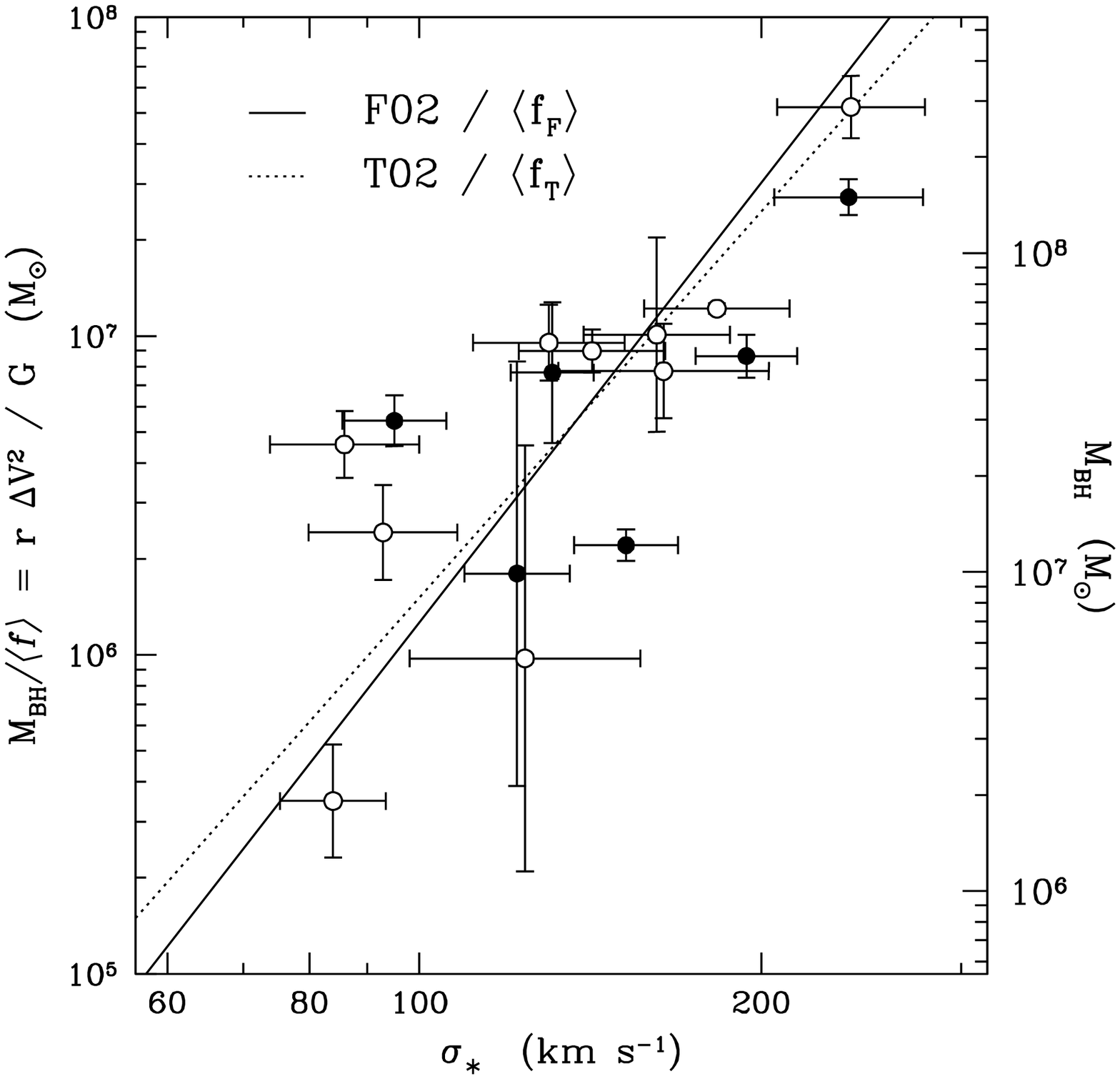} \figcaption[f2.eps]{Virial product, $M_{\rm BH}/f$, versus host
galaxy velocity dispersion, $\sigma_{\ast}$. Solid points indicate
objects with $\sigma_{\ast}$ measurements presented here. Open points
represent AGNs with previously published $\sigma_{\ast}$ data. The solid
line indicates the F02 slope of 4.58, with the y-intercept shifted
downward by $\langle f_{\rm F}\rangle$. The dotted line denotes the T02
slope of 4.02, with the y-intercept scaled down by $\langle f_{\rm
T}\rangle$. The vertical scale on the right uses our derived offset:
$\langle f_{\rm F}\rangle$=$\langle f_{\rm T}\rangle$=5.5. \label{fig2}}

\clearpage

\plotone{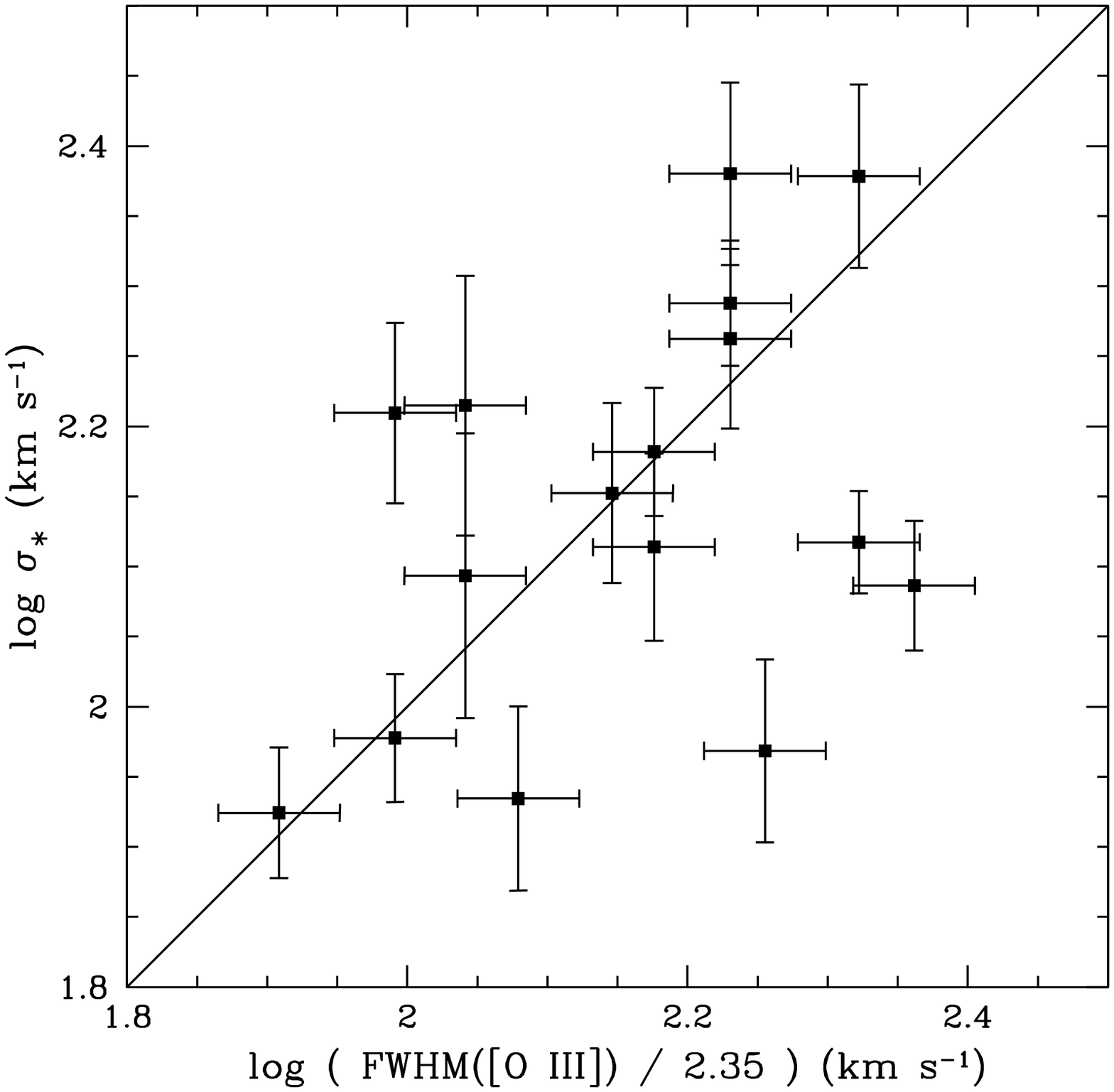} \figcaption[f3.eps]{$\sigma_{\ast}$ versus FWHM([\ion{O}{3}])
for our 16 AGNs. The solid line
shows a 1:1 correspondence between $\sigma_{\ast}$ and the equivalent for 
a Gaussian line profile, FWHM([\ion{O}{3}]) / 2.35. \label{fig3}}

\clearpage

\begin{deluxetable}{llllcl}
\tabletypesize{\footnotesize}
\tablecaption{Observing Log \label{tab1}}
\tablewidth{0pt}
\tablehead{
\colhead{} & \colhead{} & \colhead{} & \colhead{} & \colhead{Resolution} & \colhead{Targets} \\
\colhead{Run} & \colhead{Telescope} & \colhead{Instrument} & \colhead{UT Dates} & 
\colhead{(km s$^{-1}$)} & \colhead{Observed}
}
\startdata
MDM-a & MDM 2.4m & ModSpec & 2001 Oct 20--24 & 95 & Mrk 590, NGC 3227, NGC 7469 \\
KPNO & KPNO 4m & R-C Spec & 2001 Oct 29--Nov 1 & 60 & Akn 120, Mrk 590, NGC 7469 \\
MDM-b & MDM 2.4m & ModSpec & 2003 Mar 12--13 & 95 & IC 4329A, NGC 3783 \\
CTIO & CTIO 4m & R-C Spec & 2003 Apr 18 & 90 & IC 4329A, NGC 3783 \\
\enddata
\end{deluxetable}

\begin{deluxetable}{llcll}
\tabletypesize{\footnotesize}
\tablecaption{Velocity Dispersion Measurements \label{tab2}}
\tablewidth{0pt}
\tablehead{
\colhead{} & \colhead{} & \colhead{Redshift} & \colhead{} & \colhead{$\sigma_{\ast}$}\\
\colhead{Galaxy} & \colhead{Redshift} & \colhead{References} & \colhead{Run} & \colhead{(km s$^{-1}$)}
}
\startdata
Akn 120 & 0.032296$\pm$0.000143 & 1 & KPNO & 239$\pm$36 \\
IC 4329A & 0.016054$\pm$0.000050 & 2 & CTIO & 131$^{+20}_{-60}$ \\
\nodata &  & & MDM-b & 121$\pm$18 \\
Mrk 590 & 0.026385$\pm$0.000040 & 3 & KPNO & 201$\pm$30 \\
\nodata & & & MDM-a & 188$\pm$28 \\
NGC 3227 & 0.003859$\pm$0.000010 & 3 & MDM-a & 139$\pm$21 \\
NGC 3783 & 0.009730$\pm$0.000007 & 4 & CTIO & \phantom{1}87$\pm$13 \\
\nodata & & & MDM-b & 108$\pm$16 \\
NGC 7469 & 0.016317$\pm$0.000007 & 5 & KPNO & 149$\pm$22 \\
\nodata & & & MDM-a & 157$\pm$24 \\
\enddata
\tablerefs{(1) Falco et al.\ 1999; (2) Willmer et al.\ 1991; (3) de Vaucouleurs et al. 1991;
(4) Theureau et al.\ 1998; (5) Keel 1996.}
\end{deluxetable}

\begin{deluxetable}{llllc}
\tabletypesize{\footnotesize}
\tablecaption{M$_{\rm BH}$--$\sigma_{\ast}$ Data \label{tab3}}
\tablewidth{0pt}
\tablehead{
\colhead{} & \colhead{Virial Product\tablenotemark{a}} & \colhead{Black Hole Mass\tablenotemark{b}} & \colhead{$\sigma_{\ast}$ (avg)} & \colhead{$\sigma_{\ast}$}\\
\colhead{Galaxy} & \colhead{(10$^{6}$ M$_{\odot}$)} & \colhead{(10$^{6}$ M$_{\odot}$)}  & \colhead{(km s$^{-1}$)} & \colhead{References}
}
\startdata
3C 120 & \phantom{1}10.1$^{+5.7}_{-4.1}$ & 55.6$^{+31.4}_{-22.3}$ & 162$\pm$24 & 2 \\
3C 390.3 & \phantom{1}52.2$\pm$11.7 & \phantom{.}289$\pm$64 & 240$\pm$36 & 3 \\
Akn 120 &\phantom{1}27.2$\pm$3.5 & \phantom{.}150$\pm$19 & 239$\pm$36 & 1 \\
IC 4329A\tablenotemark{c} & \phantom{1}1.80$^{+3.25}_{-2.16}$ & \phantom{1}9.9$^{+17.9}_{-11.9}$ & 122$\pm$13 & 1 \\
Mrk 79 & \phantom{1}9.52$\pm$2.61 & 52.4$\pm$14.4 & 130$\pm$20 & 4 \\
Mrk 110 & \phantom{1}4.57$\pm$1.10 & 25.1$\pm$6.1 & \phantom{1}86$\pm$13 & 4 \\
Mrk 590 & \phantom{1}8.64$\pm$1.34 & 47.5$\pm$7.4 & 194$\pm$20 & 1 \\
Mrk 817 & \phantom{1}8.98$\pm$1.40 & 49.4$\pm$7.7 & 142$\pm$21 & 4 \\
NGC 3227 & \phantom{1}7.67$\pm$3.90 & 42.2$\pm$21.5 & 131$\pm$11 & 1, 2 \\
NGC 3516 & \phantom{1}7.76$\pm$2.65 & 42.7$\pm$14.6 & 164$\pm$35 & 5\\
NGC 3783 & \phantom{1}5.42$\pm$0.99 & 29.8$\pm$5.4 & \phantom{1}95$\pm$10 & 1 \\
NGC 4051 & 0.348$\pm$0.142 & 1.91$\pm$0.78 & \phantom{1}84$\pm$9 & 2, 4 \\
NGC 4151 & \phantom{1}2.42$\pm$0.83 & 13.3$\pm$4.6 & \phantom{1}93$\pm$14 & 4 \\
NGC 4593\tablenotemark{c} & \phantom{1}0.98$^{+1.70}_{-1.26}$ & 5.36$^{+9.37}_{-6.95}$ & 124$\pm$29 & 2 \\
NGC 5548 & 12.20$\pm$0.47 & 67.1$\pm$2.6 & 183$\pm$27 & 4 \\
NGC 7469 & \phantom{1}2.21$\pm$0.25 & 12.2$\pm$1.4 & 152$\pm$16 & 1 \\
\enddata
\tablenotetext{a}{From Peterson et al.\ (2004).}
\tablenotetext{b}{Scaled using $f$ = 5.5.}
\tablenotetext{c}{Excluded from fits.}
\tablerefs{(1) This work; (2) Nelson \& Whittle 1995; (3) Green et al.\ 2003; (4) Paper I;
(5) Arribas et al.\ 1997.}
\end{deluxetable}

\begin{deluxetable}{lllll}
\tabletypesize{\footnotesize}
\tablecaption{M$_{\rm BH}$--$\sigma_{\ast}$ Fitting Results \label{tab4}}
\tablewidth{0pt}
\tablehead{
\colhead{Constraint} & \colhead{Slope ($\beta$)} & \colhead{Intercept ($\alpha$)} & 
\colhead{$\chi_{\nu}^2$} & \colhead{$\langle f\rangle$}
}
\startdata
F02 slope & 4.58 & 7.48$\pm$0.13 & 2.90 & 5.5$\pm$1.9 \\
T02 slope & 4.02 & 7.39$\pm$0.12 & 2.87 & 5.5$\pm$1.7 \\
none & 4.11$\pm$1.07 & 7.40$\pm$0.21 & 3.11 & N/A \\
F02 slope, no NGC~4051 & 4.58 & 7.50$\pm$0.14 & 3.09 & 5.2$\pm$1.9 \\
T02 slope, no NGC~4051 & 4.02 & 7.42$\pm$0.12 & 2.96 & 5.1$\pm$1.6 \\
\enddata
\end{deluxetable}

\end{document}